\documentclass[sigconf]{acmart} 

\AtBeginDocument{%
  }

\setcopyright{none}
\renewcommand\footnotetextcopyrightpermission[1]{}
\copyrightyear{2026}
\acmYear{2026}
\setcopyright{cc}
\setcctype{by}
\acmConference[CHI EA '26]{Extended Abstracts of the 2026 CHI Conference on Human Factors in Computing Systems}{April 13--17, 2026}{Barcelona, Spain}
\acmBooktitle{Extended Abstracts of the 2026 CHI Conference on Human Factors in Computing Systems (CHI EA '26), April 13--17, 2026, Barcelona, Spain}
\acmDOI{10.1145/3772363.3798766}
\acmISBN{979-8-4007-2281-3/2026/04}

\usepackage{graphicx}
\usepackage{tikz}
\usepackage{color}
\usepackage{subcaption}
\usepackage{booktabs}
\usepackage{xcolor}
\usepackage{setspace}
\usepackage{tabularx}        
\usepackage{enumitem}

\definecolor{mypurple}{RGB}{119, 69, 198}

\begin{document}

\title{Co-designing for the Triad: Design Considerations for Collaborative Decision-Making Technologies in Pediatric Chronic Care}





\author{Ray-Yuan Chung}
\email{raychung@uw.edu}
\affiliation{%
  \institution{University of Washington}
  \city{Seattle, WA}
    \country{USA}
}

\author{Jaime Snyder}
\email{jas1208@uw.edu}
\affiliation{%
  \institution{University of Washington}
  \city{Seattle, WA}
    \country{USA}
}

\author{Zixuan Xu}
\email{zinaxu7@uw.edu}
\affiliation{%
  \institution{University of Washington}
  \city{Seattle, WA}
    \country{USA}
}

\author{Daeun Yoo}
\email{daeunyoo@uw.edu}
\affiliation{%
  \institution{University of Washington}
  \city{Seattle, WA}
    \country{USA}
}

\author{Athena C. Ortega}
\email{athena.ortega@seattlechildrens.org}
\affiliation{%
  \institution{University of Washington}
  \city{Seattle, WA}
    \country{USA}
}

\author{Wanda Pratt}
\email{wpratt@uw.edu}
\affiliation{%
  \institution{University of Washington}
  \city{Seattle, WA}
    \country{USA}
}

\author{Aaron Wightman}
\email{Aaron.Wightman@seattlechildrens.org}
\affiliation{%
  \institution{University of Washington}
  \city{Seattle, WA}
    \country{USA}
}

\author{Ryan Hutson}
\email{rhutson1@jhmi.edu}
\affiliation{%
  \institution{Johns Hopkins University}
  \city{Baltimore, MD}
    \country{USA}
}

\author{Cozumel Pruette}
\email{csouthe1@jhmi.edu}
\affiliation{%
  \institution{Johns Hopkins University}
  \city{Baltimore, MD}
    \country{USA}
}

\author{Ari Pollack}
\email{ari.pollack@seattlechildrens.org}
\affiliation{%
  \institution{Seattle Children’s Hospital}
  \city{Seattle, WA}
    \country{USA}
}

\renewcommand{\shortauthors}{Ray-Yuan Chung et al.}

\begin{abstract}
In pediatric chronic care, the triadic relationship among patients, caregivers, and healthcare providers introduces unique challenges for youth in managing their conditions. Diverging values, roles, and asymmetrical situational awareness across decision-maker groups often hinder collaboration and affect health outcomes, highlighting the need to support collaborative decision-making. We conducted co-design workshops with 6 youth with chronic kidney disease, 6 caregivers, and 7 healthcare providers to explore how digital technologies can be designed to support collaborative decision-making. Findings identify barriers across all levels of situational awareness, ranging from individual cognitive and emotional constraints, misaligned mental models, to relational conflicts regarding care goals. We propose design implications that support continuous decision-making practice, align mental models, balance caregiver support and youth autonomy development, and surface potential care challenges. This work advances the design of collaborative decision-making technologies that promote shared understanding and empower families in pediatric chronic care.


\end{abstract}






\maketitle
\section{Introduction}
Collaborative decision-making is essential in pediatric chronic care, where it is beneficial for youth patients, family caregivers, and healthcare providers (HCP) to work together to make informed treatment choices \cite{Miller_2018,Stacey_Legare_2020}. However, achieving this collaboration is difficult because decision-makers often hold divergent goals, priorities, and needs, which can create tensions and misalignment  \cite{Boland_Graham_2019,Bosch_Siebel_Heiser_Inhestern_2025,Miller_2018}. With 40\% of U.S. school-aged youth living with at least one chronic condition \cite{CDC_2024}, addressing ineffective collaborative decision-making is a pressing challenge that directly impacts short- and long-term health consequences \cite{Tong_Morton_Howard_McTaggart_Craig_2011}. Researchers in human-computer interaction (HCI) have increasingly recognized the importance of designing patient-centered decision-support systems that reflect the complex realities of living with chronic illness \cite{Dunbar_Bascom_Pratt_Snyder_Smith_Pollack_2022,Dunbar_Pratt_Bascom_Currier_Garcia_Smith_Snyder_Pollack_2024,Jacobs_Clawson_Mynatt_2016,Jacobs_Clawson_Mynatt_2014,Nikkhah_John_Yalamarti_Mueller_Miller_2022}. However, existing technologies are not suitable to support collaborative decision-making in pediatric chronic care because most focus on dyadic relationships (e.g., adult patient–provider) \cite{Guo_Xiao_Liu_Chen_Tong_Liu_2025,Hsu_Lau_Huang_Ghiloni_Le_Gilroy_Abrahamson_Moore_2016}, overlooking the need to synchronize priorities, establish shared understanding, and align treatment goals across multiple groups.

We explore these challenges through the lens of youth diagnosed with chronic kidney disease (CKD), a life-threatening condition characterized by medical complexity and intensive treatment requirements \cite{NIDDK_KidneyDiseaseChildren}. We draw on Endsley’s team situational awareness framework \cite{Endsley_2016} to understand how decision-makers coordinate effectively when they maintain a shared understanding of goals, roles, and context. Situational awareness is defined as a three level framework: \textit{“the perception of elements in the environment within a volume of time and space (Level 1), the comprehension of their meaning (Level 2), and the projection of their status in the near future (Level 3)”} \cite{Endsley_1988_SAEnhancement}. In health contexts, breakdowns in situational awareness can result in miscommunication, undermine collective decision-making, and ultimately lead to poorer patient outcomes and experiences \cite{Pollack_Mishra_Apodaca_Khelifi_Haldar_Pratt_2020, Kusunoki_Sarcevic_Zhang_Yala_2015}. Using the case of youth with CKD, we aim to demonstrate how situational awareness informs the understanding of triadic care relationships and expand the conceptual tools available for designing collaborative decision-making technologies. We explore the following research questions:
\begin{enumerate}
    \item RQ 1: What barriers do patients, caregivers, and HCPs identify across situational awareness levels that impact collaborative decision-making in pediatric CKD care?
    \item RQ 2: How can technologies be designed to support situational awareness needs among patients, caregivers, and HCPs to enable effective collaborative decision-making?
\end{enumerate}

To address these questions, we conducted workshops with youth with CKD, caregivers, and pediatric HCPs to co-design speculative collaborative decision-making technologies. Together, we generated design concepts and iteratively refined features. This study contributes to HCI by (1) identifying barriers to collaborative decision-making in pediatric chronic care, and (2) proposing design guidelines for collaborative decision-making technologies that enhance unmet situational awareness needs across decision-makers.

\section{Related Work}

\subsection{Collaborative Decision-making in Pediatric Care}
We define \textbf{\textit{collaborative decision-making}} in pediatric care as \textit{“the engagement of youth in healthcare decision-making processes while adapting to their evolving cognitive and emotional capacities”} \cite{Miller_2018}. Collaborative decision-making emphasizes inclusive and equal participation, involving youth patients in age-appropriate ways and respecting their experiences and perspectives \cite{Katz_Webb_COMMITTEE_ON_BIOETHICS_Macauley_Mercurio_Moon_Okun_Opel_Statter_2016}. Despite increased recognition of youth patients’ roles, current interventions aimed at supporting collaborative decision-making have yielded inconsistent outcomes \cite{Bosch_Siebel_Heiser_Inhestern_2025,Wyatt_List_Brinkman_PrutskyLopez_Asi_Erwin_Wang_DomecqGarces_Montori_LeBlanc_2015}. Existing tools and programs often overemphasize parental perspectives \cite{Bosch_Siebel_Heiser_Inhestern_2025} and provide limited insights into patient-caregiver-clinician dynamics \cite{Boland_Graham_2019,Wyatt_List_Brinkman_PrutskyLopez_Asi_Erwin_Wang_DomecqGarces_Montori_LeBlanc_2015}. 

\subsection{Decision-support Tools in Pediatric Care}
In HCI, CSCW, and health informatics, there is growing recognition of the need to support patients and caregivers in making value-driven decisions throughout ongoing care \cite{Cha_Saxena_Wou_Lee_Newman_Park_2022,Hong_Feustel_Agnihotri_Silverman_Simoneaux_Wilcox_2017,Hong_Wilcox_Machado_Olson_Simoneaux_2016,Park_Chen_2017,Seo_Kim_Kim_Fan_Ackerman_Choi_Park_2025,Zhu_Colgan_Reddy_Choe_2016}. However, when viewed through the lens of team situational awareness, existing decision-support tools appear fragmented, addressing only isolated requirements—such as access to information \cite{Ralston_Hirsch_Hoath_Mullen_Cheadle_Goldberg_2009}, improved comprehension through communication \cite{Hong_Lakshmi_Olson_Wilcox_2018,Seo_Kim_Kim_Fan_Ackerman_Choi_Park_2025}, or goal-setting \cite{Zhao_Kim_Apodaca_Casanova-Perez_Haldar_Mishra_Dunbar_Pollack_Pratt_2021}. Furthermore, many tools are designed for adult patients \cite{Eberhart_Slogeris_Sadreameli_Jassal_2019, Romm_Skoge_Barrett_Berentzen_Bergsager_Fugelli_Bjella_Gardsjord_Kling_Kruse_etal._2025, Hsu_Lau_Huang_Ghiloni_Le_Gilroy_Abrahamson_Moore_2016}, or in-clinic use \cite{Branco_Moteiro_etal_2024,Haldar_Khelifi_Mishra_Apodaca_Beneteau_Pollack_Pratt_2020,Weibel_Emmenegger_Lyons_Dixit_Hill_Hollan_2013, Walker_Leveille_Kriegel_Lin_Liu_Payne_Harcourt_Dong_Fitzgerald_Germak_etal._2021, Haldar_Kim_Mishra_Hartzler_Pollack_Pratt_2020,Zhao_Kim_Apodaca_Casanova-Perez_Haldar_Mishra_Dunbar_Pollack_Pratt_2021}. These solutions overlook youth who are developing decision-making capacity beyond clinical visits \cite{Dunbar_Pratt_Bascom_Currier_Garcia_Smith_Snyder_Pollack_2024}. Together, these gaps highlight the need for decision-making technologies that better support pediatric chronic care by fostering shared situational awareness among all stakeholders.


\section{Methods}

\subsection{Participants.}
We conducted a total of seven 2-hour co-design workshops via Zoom, each including a single decision-maker group, with the goal of having 2 to 4 participants per session and reducing power imbalances. This study received ethical approval from the University of Washington Institutional Review Board and Seattle Children's Hospital Ethics Committee. All participants were recruited through convenience sampling from a single large children’s hospital in the United States either in person during clinic visits, or from the patient portal. Eligibility criteria included youth aged 12–25 years with pre-dialysis CKD stages 3–5, and their caregivers. HCPs were eligible if they have experience working in pediatric nephrology. Nineteen participants were recruited, including 6 youth (mean age 15.3 years, range: 13–17), 6 caregivers, and 7 HCPs (including physicians, a nurse, a physician assistant, a dietitian, and a social worker). Demographic characteristics of participants are summarized in Appendix 1.

\subsection{Co-design Workshop Procedure.}
Workshops began by introducing the concept of collaborative decision-making and the potential role of technology in improving decision-making processes. Participants reflected on past experiences navigating medical decisions, responding to prompts tailored to their roles. Participants then completed two design activities (described below), followed by a group discussion on how their feature suggestions align with values.

\subsubsection{Vignette-based Role-Play for Values Elicitation} Participants were introduced to a visual storyboard (Appendix 2) portraying fictional characters navigating three phases of CKD treatment plans after a CKD diagnosis: an initial clinic visit, experiences outside of clinic, and a follow-up clinic visit. The goal of the storyboard was to prompt discussion around values, misaligned goals, and communication gaps by capturing the context- and time-dependent nature of patient values in relation to concrete events, similar to other uses of patient journey mapping \cite{Bui_Oberschmidt_2023} and scenario-based values elicitation \cite{Lee_Roldan_Zhu_KaurSaluja_Na_Chin_Zeng_Lee_Yip_2021}. The ecological validity of the vignette was vetted by two pediatric nephrologists and two patient advisory board members. Participants were asked to annotate the storyboard with pre-defined prompts, using digital sticky notes to identify key barriers of each character (patient, caregiver, HCP) for effective decision-making. 


\subsubsection{"Magical Tool" Ideation}
Building on insights from the storyboard, participants were invited to design a "magical tool" that could alleviate pain points identified in the previous exercise and enhance collaborative decision-making. This speculative design exercise is inspired by Iacucci et al. \cite{Iacucci_Kuutti_Ranta_2000}, to encourage open-ended, creative ideation before translating concepts into concrete system features. Participants described their “magical tool" using text boxes or sketches on Google Slides. Reflective prompts were given, such as: \textit{“How would this tool help you express your values or concerns?”}, and \textit{“What kinds of information should the tool display?”}. Appendix 3 shows a magic tool example sketched by one of the participants. 


\subsection{Data Analysis.}
Transcripts, storyboards, and magic tool prototypes were securely stored on a HIPAA-compliant shared drive. All data were anonymized using participant identifiers and decision-maker categories. A six-step mixed inductive–deductive thematic analysis \cite{Clarke_Braun_2014} was used to examine textual and visual data collected during the study, including workshop transcripts and participant-generated artifacts. In the inductive phase, four researchers created affinity diagramming \cite{research_methods_hci_2017}, organizing participant quotes excerpted from storyboards and the magic tool transcripts. Each researcher independently reviewed the quotes and then collaboratively grouped them under preliminary codes. In the deductive phase, two researchers mapped these themes onto the Situational Awareness framework \cite{Endsley_1988_SAEnhancement} refining through multiple rounds of discussion with the larger research group. 

\section{Results}
\subsection{Perception: Gathering Information about Illness, Treatment, and Options}

\subsubsection{Barriers for Perception}
HCPs described that some pediatric patients seemed detached and uninterested during visits, impacting discussions about their care. However, caregivers and patients clarified that this behavior often stems from \textbf{\textit{emotional barriers}}. Patient 5 drew on their own experience to suggest that the storyboard character’s lack of engagement likely reflects the fear and anxiety of a new diagnosis: \textit{“Assuming [the patient] just got diagnosed with [CKD], he's probably concerned with fear, anxiety and nervousness.”}. In addition, \textbf{\textit{limited time}} and \textbf{\textit{cognitive constraints}} were frequently identified as barriers to receiving relevant information during clinic visits. Both patients and caregivers emphasized the overwhelming nature of the first appointment, where large volumes of new information are presented in a short period of time. Many described struggling to retain new information while simultaneously trying to formulate questions under time pressure. As Caregiver 1 explained, \textit{“it's really hard when in the doctor's office to remember all of the things [and] to share the concerns that we had. Because oftentimes the doctor is trying to give us information that we need or ask me specific questions [about] information that they need from us.”} 


\subsubsection{Envisioning Support for Perception}
Some HCPs suggested tools to support in-clinic communication, such as artificial intelligence (AI) note-taking software that \textbf{\textit{capture key information from clinical conversations}}, freeing families to focus on communication rather than transcription. However, many patients and caregivers envisioned support beyond the clinic, such as immersive tools like Virtual Reality (VR) and gaming environments that allow them to practice conversations and rehearse clinical scenarios to \textbf{\textit{prepare for clinical interactions}}. Caregiver 2 suggested role-playing simulations: \textit{“A role play or like a simulation [patients] could play, act, talk and work with the doctor, so they would be prepared.”}. In addition, patients and caregivers alike envisioned tools that \textbf{\textit{support anxiety management and encourage self-expression}}. For example, Caregiver 3 proposed an anonymous messaging feature: \textit{“Patients could ask questions anonymously so they don't feel the pressure … because sometimes [patients] think their questions are dumb.”}. 



\subsection{Comprehension: Integrating and Interpreting Information}
\par

\subsubsection{Barriers for Comprehension}
Patients and caregivers were expected to absorb extensive treatment, medication, exercise and dietary information within a single visit, resulting in an \textbf{\textit{information overload}} that compromises their ability to process critical information. As Patient 2 shared,\textit{“There is a lot of things to process, like all the new meds you might take, the new schedules and everything … appointments can be very tiring and overwhelming”}. Compounding this challenge, patients and caregivers often struggled with \textbf{\textit{limited mental models}} about CKD and its implications, which made it difficult to interpret lifestyle options and trade-offs. Here, we define \textit{mental models} as the background knowledge and contextual understanding that shape how individuals interpret information and make sense of medical guidance, which can vary critically across individuals \cite{Evans_RossBaker_2012}. Dietary changes were a particularly frequent source of concern. While HCPs provide extensive recommendations, these guidelines often clash with families’ existing mental models of how their children eat and live, creating barriers to adherence. As Caregiver 6 mentioned: \textit{“Is there going to be like, taking away 90\% of the stuff [the patient] likes to eat or none of the stuff they like to eat? … [I'm] thinking of what to ask for types of food [the patient] needs to eat and types they needs to stay away from.”}. 

\subsubsection{Envisioning Support for Comprehension}
Participants across groups envisioned tools that \textbf{\textit{translate complex medical information into accessible, visually rich formats}} tailored to non-experts. HCP 2 reimagined patient education materials using visual media: \textit{“Maybe turning some of the patient education materials into short videos, animated or in the style of social media or TikTok, where it's approachable for teenagers.”}. Furthermore, consensus across groups highlighted a need for tools that \textbf{\textit{scaffold lifestyle change alternatives and provide personalized feedback}} to help families integrate treatment recommendations into daily life. Caregiver 2 proposed a \textit{“Diet and Lifestyle Recommendation Options Navigator”}: \textit{“You could put different foods up there and [users] could choose the good foods and recognize bad foods. And so the more good foods they chose, the more points they would get.”}. 





\subsection{Projection: Linking Treatment Plans to Impact}
\par

\subsubsection{Barriers for Projection}
Patients and caregivers faced challenges in \textbf{\textit{anticipating how treatment would affect routines and social lives}}. Patient 5 reflected on how understanding the long-term implications earlier would have influenced their behavior: \textit{“I would probably take the information more seriously ... [if I understood] listening and doing those activities that are recommended to me by my doctor would help me live longer and live a much more healthier life.”}. For some families, caregiver presence during clinic visits became a barrier to open communication, which indirectly hindered the HCP’s ability to comprehend patients' struggles and project actionable plans. These dynamics sometimes lead to \textbf{\textit{patient–caregiver conflicts}}, which surfaces as a notable barrier when patients’ concerns and caregivers’ priorities do not fully align. Some patients value having private conversations with the medical team to express their perspectives more freely, as Patient 4 expressed: \textit{“I would probably ask for a more one-on-one experience, so that I don't feel like someone else is watching. I'm about to turn [into] an adult, I have all this stuff going on, and I don't know what this is about. … as patients get older, they'll probably have more of a louder voice and care more about what they want.”}. Only a few caregivers acknowledged this tension between providing parental support and respecting patient autonomy. Caregiver 5 acknowledged the importance of creating space for patients’ voices, but noted that this transition often takes time: \textit{“I need to probably sit back and not talk as much and let [my child] speak up for themselves, because I've always spoken for them. So all of a sudden, it’s [hard to realize] they can actually speak for themselves [as a teenager].”}.

\subsubsection{Envisioning Support for Projection}
Participants across groups integrated \textbf{\textit{goal-setting and projections of potential challenges}} features in their magic tools. HCP 6 suggested a game-like environment where users could explore and prepare for upcoming treatment challenges: \textit{“A game that [users] could put their goals in to manage med management, salt counting, etc.”}. In addition, collaborative decision-making technologies should be designed for patients, caregivers, and, in some cases, HCPs to support collaboration and reduce power imbalances in decision-making. Caregivers envisioned tools that \textbf{\textit{facilitate shared ownership of care via collaborative features}} to make patient preferences visible and actionable. Caregiver 6 suggested a feedback loop for sharing preferences: \textit{“Apps could have a star [review system], like 5 stars, ‘It was good.’ And then that information could go to the parent, so then [they] know this is [a recipe] I should cook again.”}. 


\section{DISCUSSION}

\subsection{Design Considerations}
Grounded in these insights, we propose a set of design principles for future collaborative decision-making technologies in healthcare settings.

\textbf{Support decision-making practice outside-of-clinic.} Patients and caregivers struggled to access and absorb relevant information during clinic visits because of limited time, cognitive overload, and emotional barriers. Practice-based tools could allow them to prepare prior to appointments by rehearsing “if-then” treatment-related decisions for scenarios they may not have previously considered. Many technologies are designed primarily for in-clinic use \cite{Branco_Moteiro_etal_2024,Ralston_Hirsch_Hoath_Mullen_Cheadle_Goldberg_2009,Walker_Leveille_Kriegel_Lin_Liu_Payne_Harcourt_Dong_Fitzgerald_Germak_etal._2021}, offering limited support for the broader experience of navigating decisions across home, school,  social contexts and clinic visits. Prior work on augmented reality, virtual agents, and gaming systems shows potential in supporting mental barriers \cite{Mittmann_etal_2022}, address health knowledge gaps \cite{Steenstra_etal_2024}, and develop self-management skills \cite{denAkker_Klaassen_Bul_Kato_vanderBurg_diBitonto_2017}. 

\textbf{Aligning mental models through accessible visualizations.} Patients and caregivers can enter clinical encounters with misaligned mental models of CKD, hindering their ability to interpret treatment recommendations, anticipate trade-offs, and integrate clinical guidance into daily life. These challenges align with prior findings on mismatched interpretations arising from differing mental models \cite{Endsley_2016,Pollack_Mishra_Apodaca_Khelifi_Haldar_Pratt_2020}. Workshop participants across decision-maker groups suggested dashboards that visualize clinical metrics in accessible formats, tools that summarize and contextualize risks, and patient education materials redesigned as animations or social media--style content to better engage adolescents. These ideations align with prior systems that translate complex information into accessible formats, such as animated educational systems for youth \cite{Steenstra_etal_2024} and approaches that convert academic papers into design cards \cite{shin_2024}.

\textbf{Enabling shared ownership of care.} Youth patients expressed a desire to gradually assume more responsibility for their care, but many struggled to exercise agency, leading caregivers to fill the gap. This dynamic is further complicated by the caregivers' own difficulty in relinquishing decision-making authority. Mismatched expectations about when and how caregivers should participate versus when patients should have greater voice often created friction in collaborative decision-making, reflecting prior work \cite{Boland_Graham_2019,Hong_Wilcox_Machado_Olson_Simoneaux_2016}. At the same time, it is important to account for patients’ developmental stages when designing technologies to support evolving patient--caregiver dynamics \cite{Toscos_Connelly_Rogers_2012}. Insights from workshops suggest that multi–decision-maker interaction features can help align priorities and support role negotiation, echoing prior approaches such as interactive storyboards \cite{Hong_Lakshmi_Olson_Wilcox_2018} and AI chatbots \cite{Seo_Kim_Kim_Fan_Ackerman_Choi_Park_2025} that facilitate communication and shared understanding in pediatric care.

\textbf{Supporting anticipation of potential challenges}. Patients and families need support in anticipating how treatment decisions affect daily routines and social lives, where misaligned goals with HCPs often emerge. Participants across decision-maker groups envisioned technologies that emphasize goal-setting and context-aware feedback to foster shared understanding among patients, caregivers, and HCPs. Prior work supports this approach: goal-setting tools have been shown to improve alignment and collaborative planning for pediatrics \cite{Zhao_Kim_Apodaca_Casanova-Perez_Haldar_Mishra_Dunbar_Pollack_Pratt_2021}, while goal-directed features can help patients prepare for and engage more effectively in clinical interactions \cite{Schroeder_etal_2020}. 

\subsection{Future Directions}
We plan to integrate our design considerations into a refined prototype and conduct a next round of iteration involving a larger-scale, multi-site user study for evaluation. While our current work focused on CKD as a representative chronic condition, the alignment of our preliminary results with prior work \cite{denAkker_Klaassen_Bul_Kato_vanderBurg_diBitonto_2017,Hong_Wilcox_Machado_Olson_Simoneaux_2016} suggests broader transferability. We invite the HCI community to further explore how these findings might be generalized to other complex healthcare contexts, such as mental health \cite{Schuster2021, Tuijt2021} or long-term care \cite{Niedling2025, Koster2021}—where caregivers are also heavily involved in triadic decision-making—to further validate and extend the impact of our collaborative decision-making framework.

\section{CONCLUSION}
Our findings emphasize the value of designing collaborative decision-making technologies grounded in situational awareness theory and tailored to the unique dynamics of pediatric chronic care. Supporting collaborative decision-making requires more than simply providing information or helping patients interpret data. Technologies should equip users to prepare for clinical interactions, align mental models of clinical information, balance caregiver support with youth autonomy, and facilitate goal alignment and anticipation of challenges in care. By surfacing these insights, this work provides a preliminary design framework for collaborative decision-making technologies that promote shared understanding and empower families within the complex ecosystem of pediatric chronic care.

\begin{acks}

This research was supported by the National Institute of Diabetes and Digestive and Kidney Diseases (NIDDK) of the National Institutes of Health (NIH) under Award Number R01DK131134. The content is solely the responsibility of the authors and does not necessarily represent the official views of the NIH. We extend our sincere gratitude to the patients, caregivers, and healthcare providers who shared their time and expertise during our workshops. We also thank the two members of the Kids CoLab Advisory Board who reviewed study materials during pilot testing, as well as Megan Kelton, Kelsey Brace, and Stella Isarankura for their assistance with recruitment at Seattle Children’s Hospital.

\end{acks}

\balance
\bibliographystyle{ACM-Reference-Format}
\bibliography{CHI26}

@article{Koster2021,
  author  = {Koster, Luzan and Verbeek, Hilde and de Boer, Bernadette and Hamers, Jan P. H.},
  title   = {It takes three to tango: An ethnography of triadic involvement of residents, families and nurses in long-term dementia care},
  journal = {Health Expectations},
  year    = {2021},
  volume  = {24},
  number  = {4},
  pages   = {1305--1315},
  doi     = {10.1111/hex.13224}
}

@article{Niedling2025,
  author  = {Niedling, Katharina and Richter, Stefanie and H\"{a}mel, Kerstin},
  title   = {Triadic relationships in home care nursing: an integrative review of the views and experiences of older couples and nurses},
  journal = {BMC Nursing},
  year    = {2025},
  volume  = {24},
  number  = {1},
  pages   = {671},
  doi     = {10.1186/s12912-025-03378-1},
  note    = {Published 2025 Jun 27}
}

@article{Tuijt2021,
  author  = {Tuijt, R. and Rees, J. and Frost, R. and Wilcock, J. and Manthorpe, J. and Rait, G.},
  title   = {Exploring how triads of people living with dementia, carers and health care professionals function in dementia health care: A systematic qualitative review and thematic synthesis},
  journal = {Dementia (London)},
  year    = {2021},
  volume  = {20},
  number  = {3},
  pages   = {1080--1104},
  doi     = {10.1177/1471301220915068}
}

@article{Schuster2021,
  author = {Schuster, F. and Holzhüter, F. and Heres, S. and Hamann, J.},
  title = {'Triadic' shared decision making in mental health: Experiences and expectations of service users, caregivers and clinicians in Germany},
  journal = {Health Expectations},
  year = {2021},
  volume = {24},
  number = {2},
  pages = {507--515},
  doi = {10.1111/hex.13192}
}

@inproceedings{shin_2024,
  author    = {Shin, Donghoon and Wang, Lucy Lu and Hsieh, Gary},
  title     = {From Paper to Card: Transforming Design Implications with Generative AI},
  booktitle = {Proceedings of the CHI Conference on Human Factors in Computing Systems},
  series    = {CHI '24},
  year      = {2024},
  location  = {Honolulu, HI, USA},
  publisher = {Association for Computing Machinery},
  address   = {New York, NY, USA},
  articleno = {554},
  numpages  = {19},
  doi       = {10.1145/3613904.3642055}
}

@article{Schroeder_etal_2020, 
author = {Schroeder, Jessica and Karkar, Ravi and Murinova, Natalia and Fogarty, James and Munson, Sean A.}, title = {Examining Opportunities for Goal-Directed Self-Tracking to Support Chronic Condition Management}, year = {2020}, issue_date = {December 2019}, publisher = {Association for Computing Machinery}, address = {New York, NY, USA}, volume = {3}, number = {4}, url = {https://doi-org.offcampus.lib.washington.edu/10.1145/3369809}, doi = {10.1145/3369809}, journal = {Proc. ACM Interact. Mob. Wearable Ubiquitous Technol.}, month = sep, articleno = {151}, numpages = {26} }

@inproceedings{Steenstra_etal_2024, author = {Steenstra, Ian and Murali, Prasanth and Perkins, Rebecca B. and Joseph, Natalie and Paasche-Orlow, Michael K and Bickmore, Timothy}, title = {Engaging and Entertaining Adolescents in Health Education Using LLM-Generated Fantasy Narrative Games and Virtual Agents}, year = {2024}, isbn = {9798400703317}, publisher = {Association for Computing Machinery}, address = {New York, NY, USA}, url = {https://doi.org/10.1145/3613905.3650983}, doi = {10.1145/3613905.3650983}, booktitle = {Extended Abstracts of the CHI Conference on Human Factors in Computing Systems}, articleno = {126}, numpages = {8}, location = {Honolulu, HI, USA}, series = {CHI EA '24} }

@article{Mittmann_etal_2022, author = {Mittmann, Gloria and Barnard, Adam and Krammer, Ina and Martins, Diogo and Dias, Jo\~{a}o}, title = {LINA - A Social Augmented Reality Game around Mental Health, Supporting Real-world Connection and Sense of Belonging for Early Adolescents}, year = {2022}, issue_date = {October 2022}, publisher = {Association for Computing Machinery}, address = {New York, NY, USA}, volume = {6}, number = {CHI PLAY}, url = {https://doi.org/10.1145/3549505}, doi = {10.1145/3549505}, journal = {Proc. ACM Hum.-Comput. Interact.}, month = oct, articleno = {242}, numpages = {21} }

@article{Evans_RossBaker_2012, title={Shared mental models of integrated care: aligning multiple stakeholder perspectives}, volume={26}, rights={https://www.emerald.com/insight/site-policies}, ISSN={1477-7266}, DOI={10.1108/14777261211276989}, abstractNote={Purpose – Health service organizations and professionals are under increasing pressure to work together to deliver integrated patient care. A common understanding of integration strategies may facilitate the delivery of integrated care across inter-organizational and inter-professional boundaries. This paper aims to build a framework for exploring and potentially aligning multiple stakeholder perspectives of systems integration.}, number={6}, journal={Journal of Health Organization and Management}, author={Evans, Jenna M. and Ross Baker, G.}, editor={Wistow, Gerald}, year={2012}, month=oct, pages={713–736}, language={en} }

@article{Haldar_Kim_Mishra_Hartzler_Pollack_Pratt_2020, title={The Patient Advice System: A Technology Probe Study to Enable Peer Support in the Hospital}, volume={4}, ISSN={2573-0142}, DOI={10.1145/3415183}, abstractNote={SHEFALI HALDAR, University of Washington, United States YOOJUNG KIM, Seoul National University, South Korea SONALI R. MISHRA, University of Washington, United States ANDREA L. HARTZLER, University of Washington, United States ARI H POLLACK, Seattle Children’s Hospital, United States WANDA PRATT, University of Washington, United States Although peer support technologies are critical resources for patients managing health conditions, they do not address the needs of patients in the hospital (i.e., inpatients) or the unique design constraints of this healthcare setting. To examine how the design of these technologies can meet the needs of inpatients, we conducted a technology probe study with 30 pediatric and adult inpatients. We created the Patient Advice System (PAS) to enable peer support in the hospital setting, then studied how participants used and perceived it during their stay. Inpatients used the PAS to exchange emotional support and share peer advice on a range of topics (e.g., adjusting to the hospital, communicating with providers). They identified several benefits (e.g., fostered connections) and challenges (e.g., competing clinical priorities) with using the PAS in the real-world context of their hospital stay. Based on our findings, we discuss three design opportunities—highlighting local expertise, designing for dynamic engagement, and providing alternative modes of peer support—for future peer support technologies to empower inpatients and overcome the difficulties they face within the hospital. CCS Concepts: • Human-centered computing → Empirical studies in collaborative and social computing; • Applied computing → Consumer health.}, number={CSCW2}, journal={Proceedings of the ACM on Human-Computer Interaction}, author={Haldar, Shefali and Kim, Yoojung and Mishra, Sonali R. and Hartzler, Andrea L. and Pollack, Ari H. and Pratt, Wanda}, year={2020}, month=oct, pages={1–23}, language={en} }

@article{Haldar_Khelifi_Mishra_Apodaca_Beneteau_Pollack_Pratt_2020,
  title   = {Designing Inpatient Portals to Support Patient Agency and Dynamic Hospital Experiences},
  author  = {Haldar, Shefali and Khelifi, Maher and Mishra, Sonali R. and Apodaca, Calvin and Beneteau, Erin and Pollack, Ari H. and Pratt, Wanda},
  journal = {Journal of the American Medical Informatics Association},
  volume  = {27},
  number  = {12},
  pages   = {1850--1861},
  year    = {2020},
  doi     = {10.1093/jamia/ocaa209},
  url     = {https://doi.org/10.1093/jamia/ocaa209},
  language= {en}
}

@inproceedings{Cha_Saxena_Wou_Lee_Newman_Park_2022, address={New York, NY, USA}, series={CHI ’22}, title={Transitioning Toward Independence: Enhancing Collaborative Self-Management of Children with Type 1 Diabetes}, ISBN={978-1-4503-9157-3}, url={https://dl.acm.org/doi/10.1145/3491102.3502055}, DOI={10.1145/3491102.3502055}, abstractNote={Although child participation is required for successful Type 1 Diabetes (T1D) management, it is challenging because the child’s young age and immaturity make it difficult to perform self-care. Thus, parental caregivers are expected to be heavily involved in their child’s everyday illness management. Our study aims to investigate how children and parents collaborate to manage T1D and examine how the children become more independent in their self-management through the support of their parents. Through semi-structured interviews with children with T1D and their parents (N=41), our study showed that children’s knowledge of illness management and motivation for self-care were crucial for their transition towards independence. Based on these two factors, we identified four types of children’s collaboration (i.e., dependent, resistant, eager, and independent) and parents’ strategies for supporting their children’s independence. We suggest design implications for technologies to support collaborative care by improving children’s transition to independent illness management.}, booktitle={Proceedings of the 2022 CHI Conference on Human Factors in Computing Systems}, publisher={Association for Computing Machinery}, author={Cha, Yoon Jeong and Saxena, Arpita and Wou, Alice and Lee, Joyce and Newman, Mark W and Park, Sun Young}, year={2022}, month=apr, pages={1–17}, collection={CHI ’22} }

@article{Kusunoki_Sarcevic_Zhang_Yala_2015, title={Sketching Awareness: A Participatory Study to Elicit Designs for Supporting Ad Hoc Emergency Medical Teamwork}, volume={24}, ISSN={0925-9724, 1573-7551}, DOI={10.1007/s10606-014-9210-5}, abstractNote={Prior CSCW research on awareness in clinical settings has mostly focused on higher-level team coordination spanning across longer-term trajectories at the department and inter-department levels. In this paper, we offer a perspective on what awareness means within the context of an ad hoc, time- and safety-critical medical setting by looking at teams treating severely ill patients with urgent needs. We report findings from four participatory design workshops conducted with emergency medicine clinicians at two regional emergency departments. Workshops were developed to elicit design ideas for information displays that support awareness in emergency medical situations. Through analysis of discussions and clinicians’ sketches of information displays, we identified five features of teamwork that can be used as a foundation for supporting awareness from the perspective of clinicians. Based on these findings, we contribute rich descriptions of four facets of awareness that teams manage during emergency medical situations: team member awareness, elapsed time awareness, teamwork-oriented and patient-driven task awareness, and overall progress awareness. We then discuss these four awareness types in relation to awareness facets found in the CSCW literature.}, number={1}, journal={Computer Supported Cooperative Work (CSCW)}, author={Kusunoki, Diana and Sarcevic, Aleksandra and Zhang, Zhan and Yala, Maria}, year={2015}, month=feb, pages={1–38}, language={en} }

@article{Endsley_1988_SAEnhancement,
  author    = {Mica R. Endsley},
  title     = {Design and Evaluation for Situation Awareness Enhancement},
  journal   = {Proceedings of the Human Factors Society Annual Meeting},
  volume    = {32},
  number    = {2},
  pages     = {97--101},
  year      = {1988},
  doi       = {10.1177/154193128803200221},
  url       = {https://doi.org/10.1177/154193128803200221}
}

@article{Guo_Xiao_Liu_Chen_Tong_Liu_2025, title={Enhancing Doctor-Patient Shared Decision-Making: Design of a Novel Collaborative Decision Description Language}, volume={27}, ISSN={1438-8871}, DOI={10.2196/55341}, abstractNote={Background: Effective shared decision-making between patients and physicians is crucial for enhancing health care quality and reducing medical errors. The literature shows that the absence of effective methods to facilitate shared decision-making can result in poor patient engagement and unfavorable decision outcomes. Objective: In this paper, we propose a Collaborative Decision Description Language (CoDeL) to model shared decision-making between patients and physicians, offering a theoretical foundation for studying various shared decision scenarios. Methods: CoDeL is based on an extension of the interaction protocol language of Lightweight Social Calculus. The language utilizes speech acts to represent the attitudes of shared decision-makers toward decision propositions, as well as their semantic relationships within dialogues. It supports interactive argumentation among decision makers by embedding clinical evidence into each segment of decision protocols. Furthermore, CoDeL enables personalized decision-making, allowing for the demonstration of characteristics such as persistence, critical thinking, and openness. Results: The feasibility of the approach is demonstrated through a case study of shared decision-making in the disease domain of atrial fibrillation. Our experimental results show that integrating the proposed language with GPT can further enhance its capabilities in interactive decision-making, improving interpretability. Conclusions: The proposed novel CoDeL can enhance doctor-patient shared decision-making in a rational, personalized, and interpretable manner.}, journal={Journal of Medical Internet Research}, author={Guo, XiaoRui and Xiao, Liang and Liu, Xinyu and Chen, Jianxia and Tong, Zefang and Liu, Ziji}, year={2025}, month=mar, pages={e55341}, language={en} }

@article{Bosch_Siebel_Heiser_Inhestern_2025, title={Decision-making for children and adolescents: a scoping review of interventions increasing participation in decision-making}, volume={97}, rights={2024 The Author(s)}, ISSN={1530-0447}, DOI={10.1038/s41390-024-03509-5}, abstractNote={To review and synthesize the literature on interventions to facilitate shared decision-making or to increase participation in decision-making in pediatrics focusing on interventions for children and adolescents.}, number={6}, journal={Pediatric Research}, publisher={Nature Publishing Group}, author={Bosch, Inga and Siebel, Hermann and Heiser, Maike and Inhestern, Laura}, year={2025}, month=may, pages={1840–1854}, language={en} }

@inproceedings{denAkker_Klaassen_Bul_Kato_vanderBurg_diBitonto_2017,
  author    = {den Akker, Rieks op and Klaassen, Randy and Bul, Kim and Kato, Pamela M. and van der Burg, Gert-Jan and di Bitonto, Pierpaulo},
  title     = {Let Them Play: Experiences in the Wild with a Gamification and Coaching System for Young Diabetes Patients},
  booktitle = {Proceedings of the 11th EAI International Conference on Pervasive Computing Technologies for Healthcare (PervasiveHealth ’17)},
  year      = {2017},
  month     = may,
  pages     = {409--418},
  publisher = {Association for Computing Machinery},
  address   = {New York, NY, USA},
  isbn      = {978-1-4503-6363-1},
  doi       = {10.1145/3154862.3154931},
  url       = {https://dl.acm.org/doi/10.1145/3154862.3154931}
}

@article{Boland_Graham_2019,
  title     = {Barriers and facilitators of pediatric shared decision-making: a systematic review},
  author    = {Boland, Laura and Graham, Ian D. and Légaré, France and Lewis, Krystina and Jull, Janet and Shephard, Allyson and Lawson, Margaret L. and Davis, Alexandra and Yameogo, Audrey and Stacey, Dawn},
  journal   = {Implementation Science},
  volume    = {14},
  number    = {1},
  pages     = {7},
  year      = {2019},
  month     = dec,
  publisher = {Springer Nature},
  doi       = {10.1186/s13012-018-0851-5},
  issn      = {1748-5908},
  language  = {en}
}

@inproceedings{Branco_Moteiro_etal_2024,
  address    = {Honolulu, HI, USA},
  title      = {Co-designing Customizable Clinical Dashboards with Multidisciplinary Teams: Bridging the Gap in Chronic Disease Care},
  ISBN       = {979-8-4007-0330-0},
  url        = {https://dl.acm.org/doi/10.1145/3613904.3642618},
  DOI        = {10.1145/3613904.3642618},
  booktitle  = {Proceedings of the CHI Conference on Human Factors in Computing Systems},
  publisher  = {ACM},
  author     = {Branco, Diogo and Móteiro, Margarida and Bouça-Machado, Raquel and Miranda, Rita and Reis, Tiago and Decoroso, Élia and Cardoso, Rita and Ramalho, Joana and Rato, Filipa and Malheiro, Joana and Miranda, Diana and Caniça, Verónica and Pona-Ferreira, Filipa and Guerreiro, Daniela and Leitão, Mariana and Braz, Alexandra Saúde and Ferreira, Joaquim J and Guerreiro, Tiago},
  year       = {2024},
  month      = may,
  pages      = {1--18},
  language   = {en}
}

@inproceedings{Bui_Oberschmidt_2023, address={Rapperswil Switzerland}, title={Patient Journey Value Mapping: Illustrating values and experiences along the patient journey to support eHealth design}, ISBN={979-8-4007-0771-1}, url={https://dl.acm.org/doi/10.1145/3603555.3603558}, DOI={10.1145/3603555.3603558}, abstractNote={This paper introduces patient journey value mapping – an approach to capture experiences, emotions and values implicated in patients’ care delivery. As patients’ values (i.e., what’s important to them in their lives) may change along their patient journeys, our approach aims to support designers to respond to patients’ changing needs in the (re)design of eHealth, by mapping patients’ values and their prioritisations over time. To substantiate the creation of the map, we propose two preceding data collection phases comprising complementary empirical methods. First, important care-related events and associated values are collected retrospectively through interviews, and in-situ through diary studies. Subsequently, the data are analysed to develop materials to elicit values and value tensions through deepening discussions in an interactive workshop based on which the maps are finalised. The approach is illustrated through discussions and reflections on its application in a case study investigating patient values in eHealth for rehabilitation care.}, booktitle={Mensch und Computer 2023}, publisher={ACM}, author={Bui, Michael and Oberschmidt, Kira and Grünloh, Christiane}, year={2023}, month = {September}, pages={49–66}, language={en} }

@misc{CDC_2024,
  author       = {{Centers for Disease Control and Prevention}},
  title        = {Managing Chronic Health Conditions},
  year         = {2024},
  month        = dec,
  howpublished = {\url{https://www.cdc.gov/school-health-conditions/chronic-conditions/index.html}},
  note         = {Accessed: 2025-09-11},
}

@inbook{Clarke_Braun_2014,
  author    = {Clarke, Victoria and Braun, Virginia},
  title     = {Thematic Analysis},
  booktitle = {Encyclopedia of Critical Psychology},
  year      = {2014},
  publisher = {Springer},
  address   = {New York, NY, USA},
  pages     = {1947--1952},
  doi       = {10.1007/978-1-4614-5583-7_311},
  isbn      = {978-1-4614-5583-7},
  language  = {en}
}

@article{Dunbar_Bascom_Pratt_Snyder_Smith_Pollack_2022, title={My Kidney Identity: Contextualizing pediatric patients and their families kidney transplant journeys}, volume={26}, ISSN={1399-3046}, DOI={10.1111/petr.14343}, abstractNote={BACKGROUND: Even though having a kidney transplant is the treatment of choice for children with kidney failure, it can cause anxiety for patients and their families resulting in decreased psychosocial functioning, adherence, and self-management. We set out to identify the information needs required to help pediatric patients and their families contextualize their posttransplant experiences as they recalibrate their understanding of normalcy throughout their transplant journey. METHODS: Participants submitted photographs related to feeling: (1) worried, (2) confident, (3) similar to peers without kidney disease, and (4) different from these peers. The photographs served as a foundation for an in-depth interview. RESULTS: Nineteen individuals (10 pediatric transplant recipients and 9 caregivers) were interviewed at a mean of 8 years posttransplant. We identified five specific themes and tensions our participants associated with recalibrating their version of “normal” throughout the transplant journey: (1) exchanging information (information consumers vs. information contributors, (2) transitional management (family management vs. self-management), (3) building confidence (worry vs. confidence), (4) telling one’s story (hiding vs. self-expression), and (5) normalizing kidney transplantation (feeling different vs. feeling similar). These five themes/tensions form one’s Kidney Identity, shift from negative to positive throughout the transplant journey, illustrating a more abstract and complex account of kidney transplantation over time. CONCLUSIONS: Having a patient view their Kidney Identity over time may support self-reflection of one’s progress posttransplant and potentially help clinicians, patients, and their caregivers identify barriers and areas where they may need more support to ensure their successful engagement in their care.}, number={7}, journal={Pediatric Transplantation}, author={Dunbar, Julia C. and Bascom, Emily and Pratt, Wanda and Snyder, Jaime and Smith, Jodi M. and Pollack, Ari H.}, year={2022}, month=nov, pages={e14343}, language={eng} }

@article{Dunbar_Pratt_Bascom_Currier_Garcia_Smith_Snyder_Pollack_2024, title={It’s About the Journey - Capturing Stories of the Fluctuating Experiences of Youth Kidney Transplant Patients}, volume={8}, ISSN={2573-0142}, DOI={10.1145/3653704}, abstractNote={JULIA C. DUNBAR, University of Washington, Information School, USA WANDA PRATT, University of Washington, Information School, USA EMILY BASCOM, University of Washington, Human Centered Design and Engineering, USA CARA CURRIER, Washington State University, Elson S Floyd School of Medicine, USA JOSEPH WILLIAM TAN GARCIA, University of Washington, Information School, USA JODI SMITH, Seattle Children’s Hospital, Division of Nephrology & University of Washington, School of Medicine, Department of Pediatrics, USA JAIME SNYDER, University of Washington, Information School, USA ARI H POLLACK, Seattle Children’s Hospital, Division of Nephrology & University of Washington, School of Medicine, Department of Pediatrics & University of Washington, Information School, USA Youth who undergo a kidney transplant can experience a fluctuation of successes and challenges throughout their chronic illness journey. Designing to capture their journey could help youth to reflect on their experiences, collaborate on their care, and be empowered to live their lives to the fullest. We interviewed 11 youth kidney transplant patients and 12 caregivers to elicit their transplant journey experiences. We found that probing participants about specific parts of their transplant journey gave them structure to tell us rich stories about their experiences. Based on our findings, we discuss informing the design of a tool to support the capturing of stories for youth with chronic illnesses. Designing such tool could help youth and their caregivers to identify barriers, support reflection, and promote self-efficacy. Youth with chronic illnesses already have to change so many aspects of their lives to accommodate their illness, however, by giving them a platform to capture their chronic illness journey, it could encourage them to take more control of their lives and better collaborate with others. CCS Concepts: • Human-centered computing → Empirical studies in collaborative and social computing; • Applied computing → Health informatics; Consumer health.}, number={CSCW1}, journal={Proceedings of the ACM on Human-Computer Interaction}, author={Dunbar, Julia C. and Pratt, Wanda and Bascom, Emily and Currier, Cara and Garcia, Joseph William Tan and Smith, Jodi and Snyder, Jaime and Pollack, Ari H}, year={2024}, month=apr, pages={1–26}, language={en} }

@article{Eberhart_Slogeris_Sadreameli_Jassal_2019, title={Using a human-centered design approach for collaborative decision-making in pediatric asthma care}, volume={170}, ISSN={00333506}, DOI={10.1016/j.puhe.2019.03.004}, abstractNote={Objectives: Human-centered design (HCD) is a qualitative methodology that empathizes with end-users and assists in formulating preferable and practical interventions. We explored the utility of HCD in improving pediatric asthma healthcare outcomes among patient and caregiver populations within an urban academic center. Study design: HCD employs a multiphase process that aims to identify the needs of end users and reframe solutions around each stakeholder’s preference patterns. Methods: Ethnographic-style observations were initiated among pediatric asthma healthcare providers, community environmental activists, local government public health ofﬁcials, households with a young child (<12 years of age) with asthma, and adolescent asthmatics. Common themes from the observations served as the basis for understanding users’ experiences and determining actionable areas of improvement within outpatient asthma care. Multistakeholder brainstorming sessions led to the emergence of three prototypes that underwent low-ﬁdelity ﬁeld testing. Results: The ﬁrst prototype elucidated caregivers’ preferred outpatient asthma support systems using a newly created visual decision-making aid. The second constructed prototype was a child-oriented asthma activity sheet that allowed children to better communicate their understanding and impact of asthma care. The ﬁnal prototype attempted to improve interactions between providers, caregivers, and children/adolescents using visual prompts to enhance empathetic and clinically-relevant dialogue. Conclusion: Engaging a diverse population of relevant stakeholders in disease processes that use design thinking yield relevant solutions with enhanced community buy-in. The prototypes are continuing to undergo iterative ﬁeld testing in local community and academic asthma care sites.}, journal={Public Health}, author={Eberhart, A. and Slogeris, B. and Sadreameli, S.C. and Jassal, M.S.}, year={2019}, month=may, pages={129–132}, language={en} }

@book{Endsley_2016, address={Boca Raton}, edition={2}, title={Designing for Situation Awareness: An Approach to User-Centered Design, Second Edition}, ISBN={978-0-429-14673-2}, DOI={10.1201/b11371}, abstractNote={Liberally illustrated with actual design examples, this book demonstrates how people acquire and interpret information and examines the factors that undermine this process. The second edition expands and updates the examples throughout to include a wider range of domains and increases the coverage of SA design principles and guidelines to include new areas of development.}, publisher={CRC Press}, author={Endsley, Mica R.}, year={2016}, month=apr }

@inproceedings{Hong_Feustel_Agnihotri_Silverman_Simoneaux_Wilcox_2017, address={New York, NY, USA}, series={CHI ’17}, title={Supporting Families in Reviewing and Communicating about Radiology Imaging Studies}, ISBN={978-1-4503-4655-9}, url={https://dl.acm.org/doi/10.1145/3025453.3025754}, DOI={10.1145/3025453.3025754}, abstractNote={Diagnostic radiology reports are increasingly being made available to patients and their family members. However, these reports are not typically comprehensible to lay recipients, impeding effective communication about report findings. In this paper, we present three studies informing the design of a prototype to foster patient-clinician communication about radiology report content. First, analysis of questions posted in online health forums helped us identify patients’ information needs. Findings from an elicitation study with seven radiologists provided necessary domain knowledge to guide prototype design. Finally, a clinical field study with 14 pediatric patients, their parents and clinicians, revealed positive responses of each stakeholder when using the prototype to interact with and discuss the patient’s current CT or MRI report and allowed us to distill three use cases: co-located communication, preparing for the consultation, and reviewing radiology data. We draw on our findings to discuss design considerations for supporting each of these use cases.}, booktitle={Proceedings of the 2017 CHI Conference on Human Factors in Computing Systems}, publisher={Association for Computing Machinery}, author={Hong, Matthew K. and Feustel, Clayton and Agnihotri, Meeshu and Silverman, Max and Simoneaux, Stephen F. and Wilcox, Lauren}, year={2017}, month=may, pages={5245–5256}, collection={CHI ’17} }

@inproceedings{Hong_Lakshmi_Olson_Wilcox_2018, address={New York, NY, USA}, series={CHI ’18}, title={Visual ODLs: Co-Designing Patient-Generated Observations of Daily Living to Support Data-Driven Conversations in Pediatric Care}, ISBN={978-1-4503-5620-6}, url={https://dl.acm.org/doi/10.1145/3173574.3174050}, DOI={10.1145/3173574.3174050}, abstractNote={Teens with complex chronic illnesses have difficulty understanding and articulating symptoms such as pain and emotional distress. Yet, symptom communication plays a central role in clinical care and illness management. To understand how design can help overcome these challenges, we created a visual library of 72 sketched illustrations, informed by the Observations of Daily Living framework along with insights from 11 clinician interviews. We utilized our library with storyboarding techniques, free-form sketching, and interviews, in co-design sessions with 13 pairs of chronically-ill teens and their parents. We found that teens depicted symptoms as being interwoven with narratives of personal and social identity. Teens and parents were enthusiastic about collaboratively-generated, interactive storyboards as a tracking and communication mechanism, and suggested three ways in which they could aid in communication and coordination with informal and formal caregivers. In this paper, we detail these findings, to guide the design of tools for symptom-tracking and incorporation of patient-generated data into pediatric care.}, booktitle={Proceedings of the 2018 CHI Conference on Human Factors in Computing Systems}, publisher={Association for Computing Machinery}, author={Hong, Matthew K. and Lakshmi, Udaya and Olson, Thomas A. and Wilcox, Lauren}, year={2018}, month=apr, pages={1–13}, collection={CHI ’18} }

@article{Hong_Wilcox_Machado_Olson_Simoneaux_2016, title={Care Partnerships: Toward Technology to Support Teens’ Participation in Their Health Care}, volume={2016}, DOI={10.1145/2858036.2858508}, abstractNote={Adolescents with complex chronic illnesses, such as cancer and blood disorders, must partner with family and clinical caregivers to navigate risky procedures with life-altering implications, burdensome symptoms and lifelong treatments. Yet, there has been little investigation into how technology can support these partnerships. We conducted 38 in-depth interviews (15 with teenage adolescents with chronic forms of cancer and blood disorders, 15 with their parents, and eight with clinical caregivers) along with nine non-participant observations of clinical consultations to better understand common challenges and needs that could be supported through design. Participants faced challenges primarily concerning: 1) teens’ limited participation in their care, 2) communicating emotionally-sensitive information, and 3) managing physical and emotional responses. We draw on these findings to propose design goals for sociotechnical systems to support teens in partnering in their care, highlighting the need for design to support gradually evolving partnerships in care.}, journal={Proceedings of the SIGCHI conference on human factors in computing systems . CHI Conference}, author={Hong, Matthew K. and Wilcox, Lauren and Machado, Daniel and Olson, Thomas A. and Simoneaux, Stephen F.}, year={2016}, month=may, pages={5337–5349} }

@inproceedings{Iacucci_Kuutti_Ranta_2000, address={New York City New York USA}, title={On the move with a magic thing: role playing in concept design of mobile services and devices}, ISBN={978-1-58113-219-9}, url={https://dl.acm.org/doi/10.1145/347642.347715}, DOI={10.1145/347642.347715}, abstractNote={Designing concepts for new mobile services and devices, poses several challenges to the design. We consider user participation as a way to address part of the challenges. We show how our effort relates to current and past research. In particular, PD (Participatory Design) has inspired us in developing two participatory techniques. The two techniques are organized around situations either staged or real where users and designers can envision and enact future scenarios: a role-playing game with toys, and SPES (Situated and Participative Enactment of Scenarios). They were developed in an industry-funded project that investigates services for the nomadic Internet user of the future. We then discuss how the techniques help in facing the design challenges.}, booktitle={Proceedings of the 3rd conference on Designing interactive systems: processes, practices, methods, and techniques}, publisher={ACM}, author={Iacucci, Giulio and Kuutti, Kari and Ranta, Mervi}, year={2000}, month=aug, pages={193–202}, language={en} }

@inproceedings{Jacobs_Clawson_Mynatt_2016,
  author    = {Jacobs, Maia and Clawson, James and Mynatt, Elizabeth D.},
  title     = {A Cancer Journey Framework: Guiding the Design of Holistic Health Technology},
  booktitle = {Proceedings of the 10th EAI International Conference on Pervasive Computing Technologies for Healthcare},
  series    = {PervasiveHealth '16},
  year      = {2016},
  publisher = {Association for Computing Machinery},
  address   = {New York, NY, USA},
  pages     = {1--8},
  numpages  = {8},
  doi       = {10.4108/eai.16-5-2016.2263333},
  isbn      = {978-1-63190-050-1}
}

\clearpage 
\appendix

\section{Participant demographics}

\begin{table}[H]
    \centering
    \caption{Participant demographics from the Co-Design Workshops.} 
    \label{tab:participant_demographics}
    \Description{A demographic table of 19 participants: 6 caregivers, 6 patients, and 7 healthcare providers. It lists IDs, ages, sex, roles, and years of experience.}
    
    \renewcommand{\arraystretch}{1.2}
    \setlength{\tabcolsep}{3pt} 
    
    \footnotesize 
    \begin{tabularx}{\linewidth}{l c c l X}
        \toprule
        \textbf{ID} & \textbf{Age} & \textbf{Sex} & \textbf{Role} & \textbf{YE in PN (Total Clinical YE)} \\
        \midrule
        C1 & 14 & F & Caregiver & 4 \\
        C2 & 17 & F & Caregiver & 10 \\
        C3 & 17 & M & Caregiver & 7 \\
        C4 & 14 & F & Caregiver & 1 \\
        C5 & 14 & F & Caregiver & 14 \\
        C6 & 16 & F & Caregiver & 16 \\
        P1 & 13 & F & Patient & 7 \\
        P2 & 14 & F & Patient & 10 \\
        P3 & 17 & F & Patient & 14 \\
        P4 & 16 & M & Patient & 16 \\
        P5 & 17 & M & Patient & 12 \\
        P6 & 15 & M & Patient & 15 \\
        H1 & — & M & HCP (Dietitian) & 6 (10) \\
        H2 & — & F & HCP (Social Worker) & 5 (15) \\
        H3 & — & F & HCP (Physician Assistant) & 8 (10) \\
        H4 & — & F & HCP (MD - Fellow) & 1 (4) \\
        H5 & — & M & HCP (MD) & 16 (16) \\
        H6 & — & F & HCP (MD) & 3 (6) \\
        H7 & — & F & HCP (Nurse) & 24 (24) \\
        \bottomrule
    \end{tabularx}

    \vspace{4pt}
    \begin{minipage}{\linewidth}
        \scriptsize 
        \raggedright
        \textbf{C} = caregivers; \textbf{P} = patients; \textbf{H} = healthcare providers; \\
        \textbf{Age} = for caregivers, the age of their youth patient (i.e., their child); for patients, the patient’s own age; \\
        \textbf{YE in PN} = for caregivers/patients, years since CKD diagnosis; for HCPs, years of experience in pediatric nephrology; \\
        \textbf{Total Clinical YE} = total years of clinical experience (clinicians only); “—” indicates not applicable.
    \end{minipage}
\end{table}

\section{Study Materials}
\setcounter{figure}{1} 
\renewcommand{\thefigure}{\arabic{figure}} 
\captionsetup[figure]{name=Appendix} 

\begin{figure}[H]
    \centering
    \Description{Storyboard vignette illustrating a 16-year-old named Alex, recently diagnosed with chronic kidney disease (CKD), navigating treatment decisions across three phases. Phase 1 shows Alex and caregiver Lisa at an initial clinic visit where treatment recommendations are discussed. Phase 2 depicts challenges outside the clinic as Alex struggles to follow lifestyle changes and maintain motivation. Phase 3 shows a follow-up visit where Alex, Lisa, and the clinician collaboratively discuss progress and explore new strategies, highlighting tensions, evolving goals, and communication gaps across settings.}
    \includegraphics[width=\linewidth]{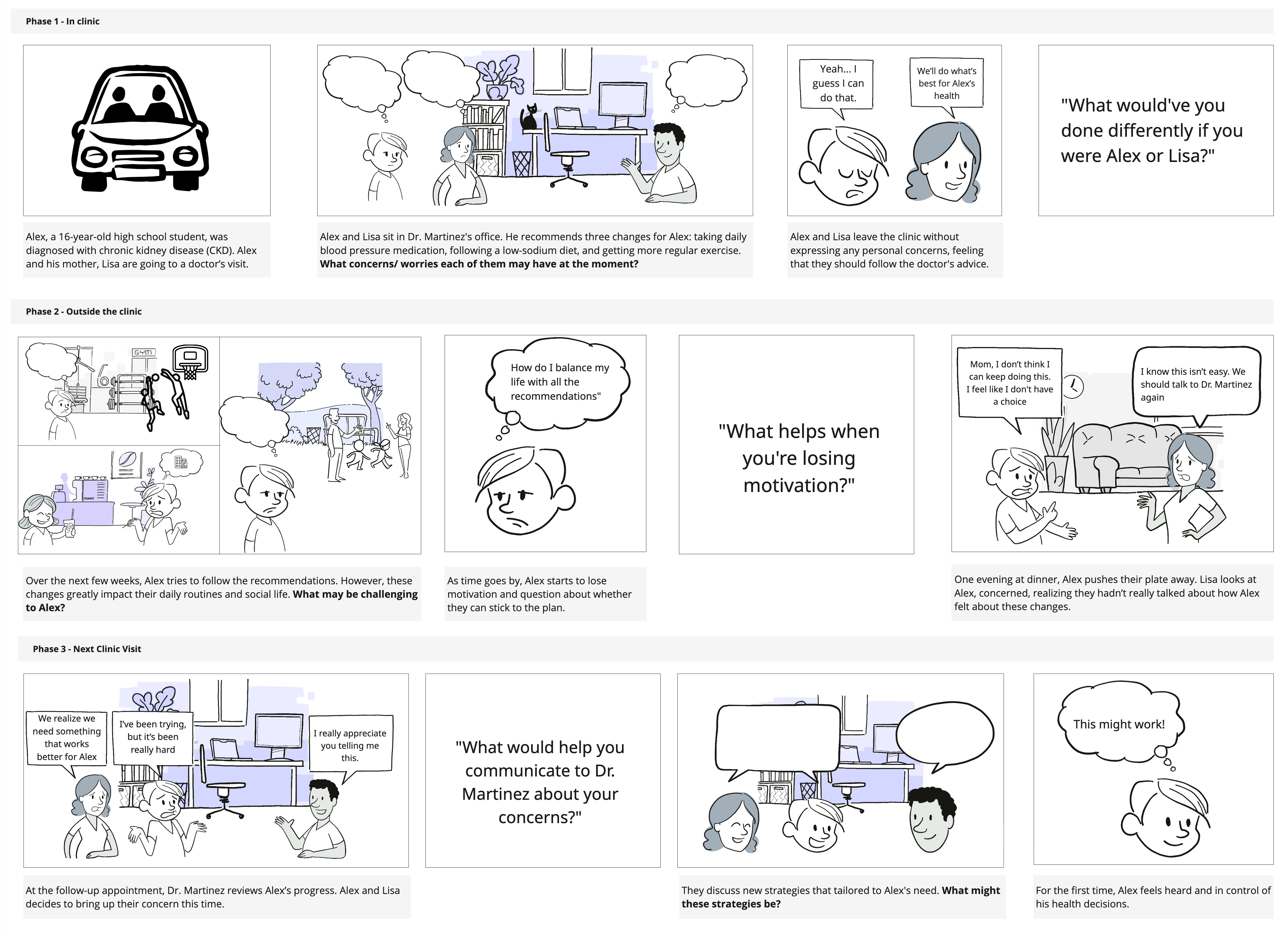} 
    \caption{Vignette storyboard used for role-playing and values elicitation during co-design sessions.}
    \label{app:storyboard}
\end{figure}

\begin{figure}[H]
    \centering
    \Description{Participant-generated sketch of a proposed “magic tool” mobile app for collaborative care planning in pediatric CKD. The concept integrates food tracking with nutritional feedback, medication and exercise monitoring, patient ratings of meals, and automatic sharing of information with caregivers and clinicians. Features include customized reward systems, visual progress charts, and goal-setting tools to support communication, engagement, and shared decision-making between patients, parents, and doctors.}
    \includegraphics[width=\linewidth]{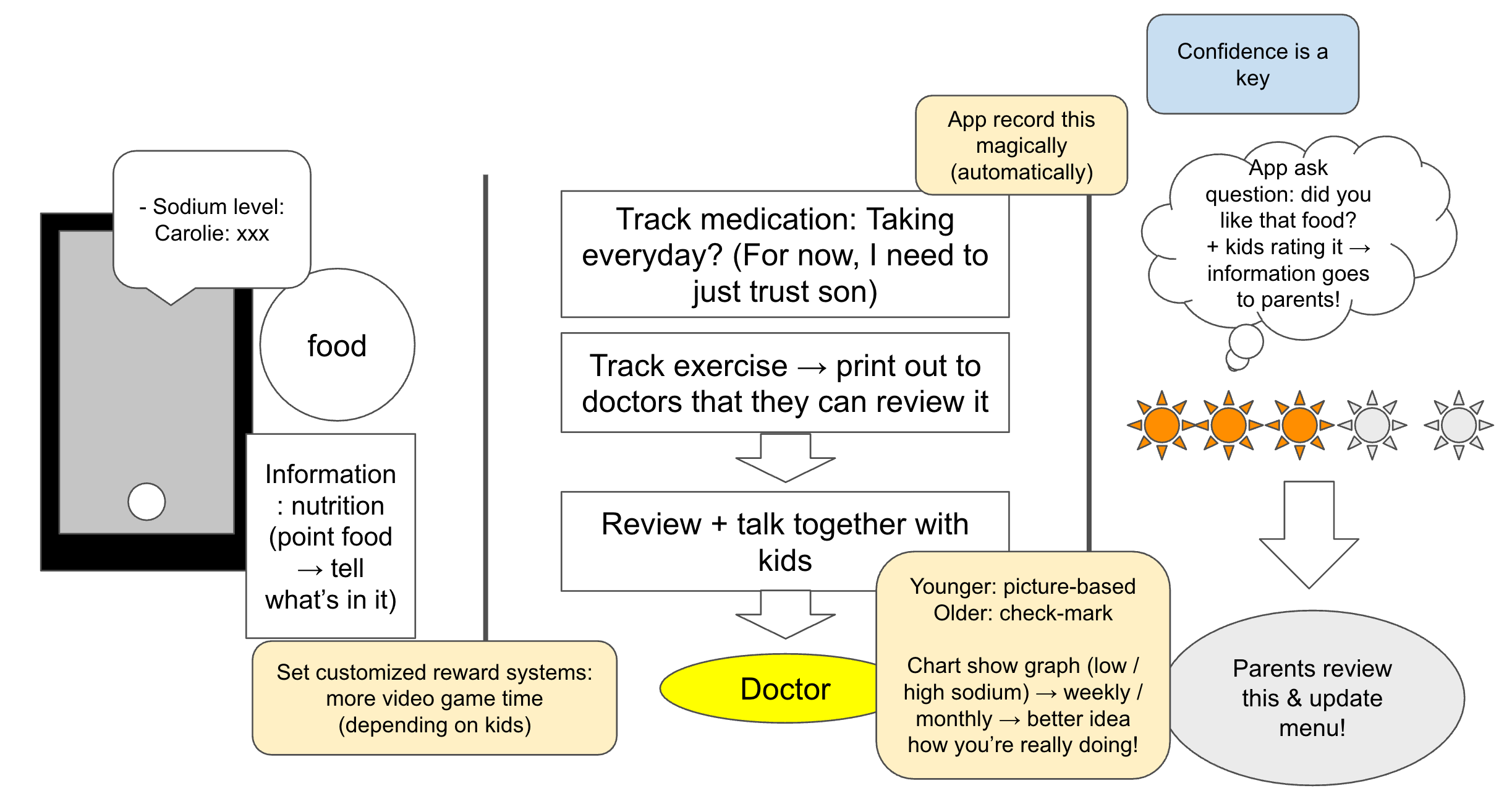} 
    \caption{Example sketch of a proposed magical tool from the co-design workshop. The tool is a mobile app with features to support gathering and contextualizing dietary information, and future planning.}
    \label{app:magictool}
\end{figure}

\end{document}